\documentstyle[mprocl,epsf]{article}
\def\Journal#1#2#3#4{{#1} {\bf #2}, #3 (#4)}

\def\PRL{\em Phys. Rev. Lett.}
\def\PRD{{\em Phys. Rev.} D}

\begin{document}

\title{OSCILLATORY AND POWER-LAW MASS INFLATION IN NON-ABELIAN BLACK HOLES}

\author{D.V. GAL'TSOV}

\address{Department of Theoretical Physics, \\
        Moscow State University, 119899 Moscow, Russia}

\author{E.E. DONETS}

\address{Laboratory of High Energies, JINR, 141980 Dubna, Russia}

\author{M.Yu. ZOTOV}

\address{Skobeltsyn Institute of Nuclear Physics, \\
        Moscow State University, 119899 Moscow, Russia}

\maketitle \abstracts{
Interior structure of non-Abelian black holes is shown to exhibit 
in a general case either an oscillating mass-inflationary
behavior, or power-law behavior with a divergent mass function. 
In both cases no Cauchy horizon forms.}

Mass inflation inside black holes emerges as back reaction
on the perturbations caused by the cross-flow of radiation tails in
the vicinity of the Cauchy horizon (CH).\cite{minfl} A particularly
simple picture of this effect arises in the case of homogeneous
(i.e., $t$--independent) perturbations in spherical black holes.
Such a situation may be also treated non-linearly as an interior problem
for a static black hole in the framework of a suitable Einstein--matter
theory. An interesting example is provided by the Einstein--Yang--Mills (EYM)
system,\cite{dgz} or its extensions including scalar fields: 
dilaton~\cite{gdz} or Higgs.\cite{gd} In the first theory an internal Cauchy
horizon may exist only for a discrete sequence of black hole
masses. For a generic mass the true CH is not formed, but, when such
a `would be' CH is approached, the mass function starts to grow
exponentially. However, the singularity is not formed instead of CH
to the contrary to the
perturbative prediction. In the full non-linear treatment this local
mass inflation stops shortly, and the metric relaxes to the next
`would be' CH. Then the process is repeated again resulting in an
oscillatory behavior of the mass function with an infinitely growing
amplitude. It is remarkable that maximal values of the mass function
attained in subsequent cycles also increase exponentially, so globally
we deal with a kind of a `quantized' mass inflation. The ultimate
singularity is spacelike and is not power-law.\cite{sing}

Spherical EYM black holes are described by a single YM function $W(r)$,
and by two metric functions $\Delta = r^2/g_{rr}$ and
$\sigma^2 = g_{tt} g_{rr}$, where $r$ is a two-sphere radius.
When $\Delta$ approaches zero (being negative) inside
the black hole,  at some $r = r_k$ the derivative $W' = dW/dr$ becomes 
approximately a linear
function of $r$, $W'= r U_k$, with an almost constant $U_k$. Then the behavior
of $\Delta$ is governed with a good accuracy by the equation
\[
      \left(\Delta/r \right)' + 2 \Delta U^2  =  0,
\]
which gives locally
\begin{equation} \label{exp}
      \Delta(r) = \frac{\Delta(r_k)}{r_k} \; r \; 
                  \exp \left [U_k^2(r^2_k-r^2) \right] .
\end{equation}
This function falls down exponentially with decreasing $r$ until
it reaches a local minimum at $R_k = 1 / (\sqrt{2} |U_k|)$.
The mass function $m(r)$ ($\Delta = r^2 - 2 m r$)
therefore is exponentially inflating when one moves
leftward from $r_k$ to $R_k$. The function $W$ stabilizes near the
limiting value although $W'$ may be very large at some tiny intervals.
The corresponding behavior of $\sigma$ follows from an approximate
integral $Z=\Delta U \sigma/r = {\rm const}$, which is
valid throughout the oscillation region:
\[
      \sigma(r) = \sigma(r_k) \exp \left[ U_k^2 (r^2 - r_k^2) \right] .
\]

After passing $R_k$, an exponential in (\ref{exp})
becomes of the order of unity,  hence $\Delta$ grows
linearly, and the mass function $m(r)$ stabilizes at the value
$M_k = m(R_k)$. Such a behavior holds until the point of the local maximum of
$\Delta/r^2$, which takes place when $\Delta \approx -V^2$ ($V=W^2-1$)
at the point
\[              
      r^*_k \approx \frac{V^2}{|\Delta(r_k)|} r_k
                    \exp \left[-(U_k r_k)^2 \right].
\]  
After that a rapid fall of $|\Delta|$ is observed causing a violent rise
of $|U|$ according to  $U \Delta \approx -V^2 U_k$, while $r$ remains
practically constant. Then $\Delta$ reaches the next local maximum at the point
$r_{k+1} \approx r_k^*$, while $m(r)$ rapidly falls down to $m_{k+1}$.

This sequence can be described by the order of magnitude
in terms of the following exponentially diverging sequence:
\[
      x_{k+1} = x_k^{-3} {\rm e}^{x_k},
\]
where $x_k = (r_k/R_k)^2 \, (\gg 1)$. In terms of $x_k$ one has
\[
      \frac{r_{k+1}}{r_k} = x_k {\rm e}^{-x_k/2}, \quad 
     |\Delta(r_k)| = x_k^{-1},
\]
so we deal with an infinite sequence of ``almost'' Cauchy horizons
as $r \rightarrow 0$. At the same time the values of  $|\Delta|$
and $m$  at $R_k$ grow rapidly
\[
     |\Delta(R_k)| = x_k^{-3/2} {\rm e}^{x_k/2},  \quad
      \frac{M_k}{M_{k-1}}=x_k^{-1}{\rm e}^{x_k/2}.
\]
Finally, $\sigma _{k+1}/ \sigma _k = {\rm e}^{-x_k/2}$.

The same theory with dilaton~\cite{gdz} predicts qualitatively
different behavior of generic black hole solutions in the interior region.
In this case after some oscillations the mass function is attracted
to a monotonic power-law solution without CH-s, which terminates in
a spacelike singularity. Similar behavior is observed in the EYM models
with doublet or triplet Higgs.\cite{gd} In both cases the attractor
solution is described by the truncated system in which kinetic
scalar terms are dominant:
\[
      \left( \ln U \right)' - 2 \phi'   =  0 , \quad
      \left[ \ln (\Delta / r) \right]'  =
      \left[ \ln (\Delta \phi'] \right)'= - r \phi'^2.
\]
Its integration gives the following five-parameter, i.e., generic
family of solutions
\begin{equation} \label{family}
      W = W_0 + b r^{2 (1 - \lambda)}, \quad
      \Delta = -2 \mu r^{(1 - \lambda^2)}, \quad
      \phi   = c + \ln \left( r^{- \lambda} \right),
\end{equation}
with constant  $W_0$, $b$, $c$, $\mu$, $\lambda$.
Parameter $\lambda$ is subject to some restrictions, which
ensure scalar dominance. They differ in the EYMD and EYMH cases.%
\cite{gdz}$^{\! , \,}$\cite{gd}
It follows from (\ref{family}) that the mass function diverges as
$r \rightarrow 0$ according to the power law $m(r) = \mu r^{-\lambda^2}$.
The corresponding $\sigma$  tends to zero as
$\sigma(r) = \sigma _1 r^{\lambda^2}$, where $\sigma _1 = {\rm const}$.

This behavior seems to be rather general for the theories, which include 
scalar fields. We term it power-law mass inflation.  No exponential
mass inflation is observed in such theories since the metric does not
approach to the internal CH at all. So, in a sense, power-law mass inflation
provides an alternative to the standard mass inflation scenario. Both
types of the mass and metric functions behavior for the EYM and EYMD
cases are shown in the figure (coordinate $r$ and all the functions are 
given in power $1/4$).

\begin{figure}[t]
\centerline{\epsfysize=8.5cm \epsfbox{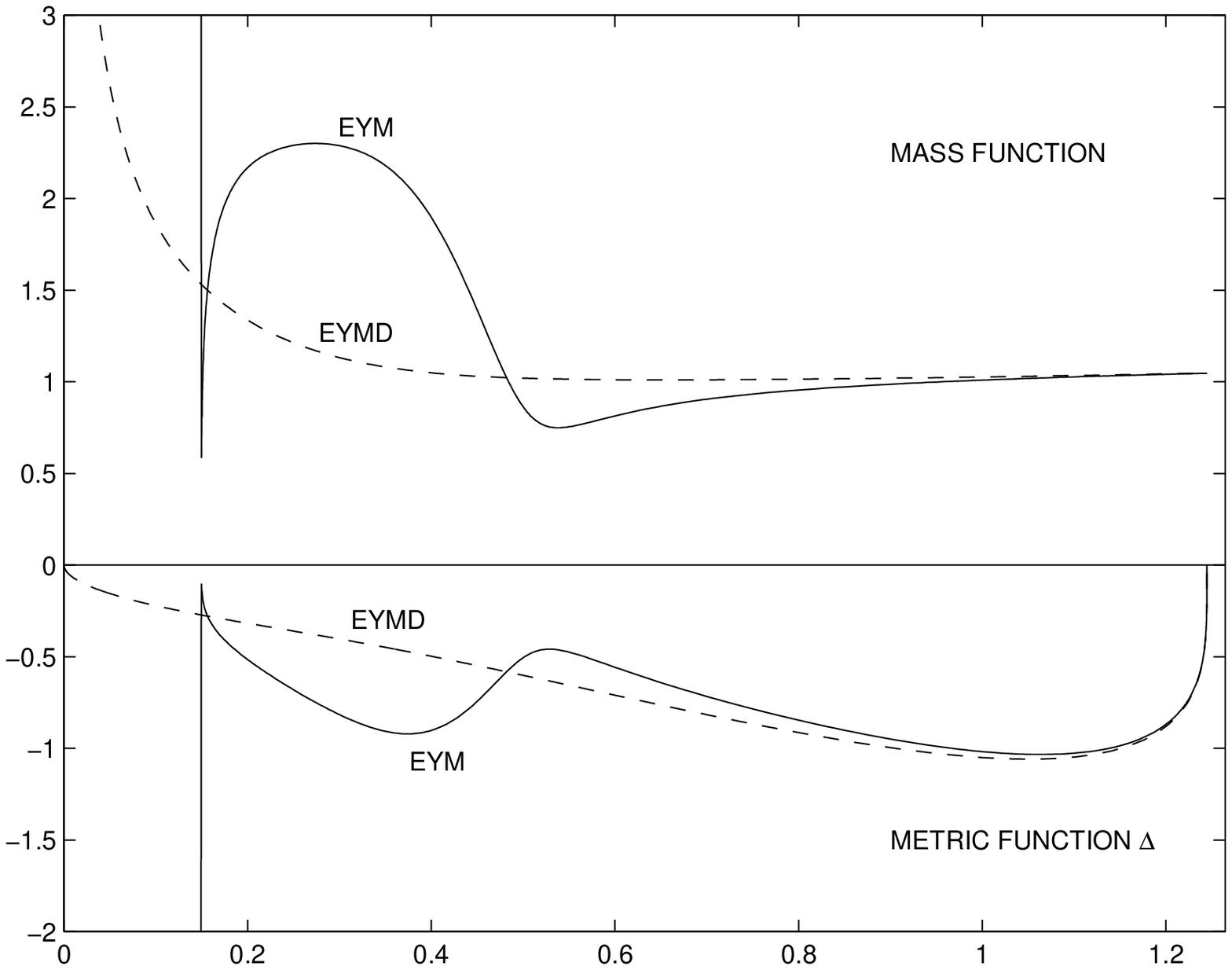}}
\end{figure}

D.V.G. is grateful to the Organizing Committee for support during the
conference. The work was supported in part by the RFBR grants 96-02-18899,
18126.

\end{document}